\documentclass[prd,preprint,amsmath,amssymb,eqsecnum]{revtex4-1}
\usepackage{graphicx}
\usepackage{dcolumn}
\usepackage{bm}

\newcommand{\intp}[1]{\int\frac{d^4{#1}}{(2\pi)^4}}

\begin{document}
\title{Quantum Gravitational Contributions to Gauge Field Theories }
\author{Yong~Tang}
\author{Yue-Liang~Wu}
\email{ytang@itp.ac.cn, ylwu@itp.ac.cn}
\affiliation{ Kavli Institute for Theoretical Physics China (KITPC)
\\ State Key Laboratory of Theoretical Physics (SKLTP) \\ Institute of
Theoretical Physics, Chinese Academy of Science, Beijing, 100190,
P.R.China}
\date{\today}

\begin{abstract}
We revisit quantum gravitational contributions to quantum gauge field theories in the gauge condition independent Vilkovisky-DeWitt formalism based on the background field method. With the advantage of Landau-DeWitt gauge, we explicitly obtain the gauge condition independent result for the quadratically divergent gravitational corrections to gauge couplings. By employing, in a general way, a scheme-independent regularization method that can preserve both gauge invariance and original divergent behavior of integrals, we show that the resulting gauge coupling is power-law running and asymptotically free. The regularization scheme dependence is clarified by comparing with results obtained by other methods. The loop regularization scheme is found to be applicable for a consistent calculation.
\end{abstract}


\maketitle


{\bf Introduction}: Quantizing general relativity is one of the most interesting and frustrating questions. Since the mass dimension of its coupling constant $\kappa=\sqrt{32\pi G}$ is negative, general relativity has isolated itself from the renormalizable quantum field theories. Nevertheless, it is undoubted that quantum gravity has effects on ordinary fields from the respective of effective theory, and may contribute to the corrections of some quantities in gauge and matter theories.

Recently, the problem that whether gravity can contribute to the running of gauge couplings has attracted much attention.  In the framework of traditional background-field method, Robinson and Wilczek(RW) \cite{RW} calculated one-loop gravitational corrections to $\beta$-function in gauge theories with naive cut-off regularization(CutR) and showed that these corrections can render all gauge theories asymptotically free by changing the gauge couplings to power-law running due to the quadratic divergences. However, this result was challenged by several authors. Following RW, it is shown in \cite{Pietrykowski} that the result obtained in \cite{RW} was gauge condition dependent, and the correction to $\beta$ function was absent in the harmonic gauge. Later, using a gauge-condition independent background-field method(Vilkovisky-DeWitt's formalism) \cite{Vilkovisky,DeWitt}, author in \cite{Toms} showed the gravitational corrections to the $\beta$ function vanished in dimensional regularization(DR) \cite{DimR}. Instead of using background-field method, the authors in \cite{Ebert} performed a diagrammatic calculation in the harmonic gauge, and found no quadratic divergences in both cut-off and dimensional regularization then vanishing correction to $\beta$-function. In \cite{TangWu}, we have checked all the calculations in the framework of diagrammatic and traditional background field methods, and demonstrated that the results are not only gauge condition dependent but also regularization scheme dependent. A new loop regularization(LORE) method \cite{YLW} then was applied to carry out the calculation. As a consequence, it was found \cite{TangWu} that there was asymptotic freedom with power-law running in the harmonic gauge condition. Later, a non-zero result was also found in \cite{Reuter} in the framework of asymptotically safe quantum gravity. Gravitational corrections to $\varphi^{4}$, Yukawa interactions and
Lee-Wick Fields are also considered \cite{WuFeng,Rodigast,Zanusso,Mackey}.

The above gravitational correction to gauge $\beta$-function is tightly connected with quadratic divergence, since the dimensional coupling $\kappa$ only picks up quadratic divergences that can lead to correction of the leading term, $\kappa^2\Lambda^2F_{\mu\nu}F^{\mu\nu}$.  In this sense, the gravity may lead to power-law running of gauge coupling by standard renormalization group analysis. In refs.\cite{Donoghue,Ellis}, the authors discussed the ambiguity of the running of gauge coupling for which we will only give some comments at the end of the paper but not try to settle down the problem, since in this note we simply present a fact that a proper regualrization scheme is crucial to handle quadratic divergence in the gauge-gravity system. Unlike the quadratic divergence in gauge theories of standard model in the flat space-time, the power counting shows that the worst divergence in the standard model is quadratic for two-point gauge Green function. The gauge invariance, however, guarantees that no quadratic divergence will survive there and the coupling is running logarithmically when a gauge symmetry-preserving regularization scheme is adopted. With gravity included, the gauge invariance may not forbid the appearance of such a quadratic divergence any more, and we show that the Slavnov-Taylor identity is preserved \cite{TangWuL} for the quadratic divergences coming from gravity's contribution.

Two important questions result from the above honest and complicated calculations but inconsistent results: how to make a gauge-condition independent calculation and how to appropriately treat quadratically divergent loop integrals. The first one is thought to be solved by using the gauge-condition independent Vilkovisky-DeWitt formalism \cite{Vilkovisky,DeWitt} in the framework of background field method, which will be briefly introduced below. Recently, both \cite{heatkernel} and \cite{he} adopted Vilkovisky-DeWitt formalism, but their results do not agree each other\cite{convention}. So, it is desirable to have another calculation.  The second one will be discussed further in this note. The crucial point is to realize a symmetry-preserving regularization scheme which can treat the quadratically divergent tensor-type integrals consistently and meanwhile maintain the divergent behaviour of original integrals.

In this note, by using the gauge-condition independent Vilkovisky-DeWitt formalism and a general relation between the regularized tensor- and scalar-type quadratically divergent integrals, we compute the quantum gravitational contributions to gauge coupling. We calculate in a general way such that the results are applicable to any specific regularization scheme. We then show how the quantum gravitational contributions to $\beta$ function of gauge couplings depend on different regularization schemes. We arrive at a conclusion that the quantum gravitational contributions will change the gauge couplings to power-law running and make all gauge theories asymptotically free. This result is universal for any regularization schemes that can  preserve both gauge invariance (through consistency conditions between tensor-type and scalar-type divergent integrals) and original divergent behaviour of integrals.  As the loop regularization method has been shown to satisfy such a requirement and consistently applied in many cases\cite{YLW,LORE}, its consistency and advantage beyond one loop order has been demonstrated in detail by merging with Bjorken-Drell's analogy between Feynman diagrams and electric circuits\cite{LORE1}. There is no doubt that the loop regularization method should be appropriately applicable to quantum gravitational contributions with a consistent result.


{\bf Vilkovisky-DeWitt effective action}: We begin with a brief introduction to the Vilkovisky-DeWitt effective action. The original idea is due to \cite{Vilkovisky,DeWitt}. The key observation is that gauge condition dependence of the effective action can be reduced to the parametrization dependence of gauge field. To remove the parametrization dependence, the concept that field space is associated with its metric is introduced and the effective action is defined accordingly such that it is invariant under reparametrization of fields. More details and pedagogical review can be found in \cite{DJToms,Tomsbook}.

DeWitt's condensed index notation \cite{DeWittbook} and Riemannian metric are used throughout the paper. Let $S[\varphi]$ represent the classical action functional, it is gauge invariant under the transformation
\begin{equation}
\delta \varphi^i = K^i_{\alpha}[\varphi]\delta \epsilon^\alpha,
\end{equation}
with $K^{i}_{\alpha}[\varphi]$ regarded as the generator of gauge transformations. To quantize gauge theory, a gauge condition has to be imposed $\chi ^\alpha [\varphi]=0$. Require $\chi^\alpha[\varphi+\delta\varphi]=\chi^\alpha[\varphi]$ hold only if $\delta\chi^\alpha=0$, one has
\begin{equation}
\chi^{\alpha}{}_{,i}[\varphi] K^{i}_{\beta}[\varphi]\delta \epsilon^{\beta} \equiv Q^{\alpha}{}_{\beta}[\varphi] \delta \epsilon^{\beta} =0 \ .
\end{equation}
Thus the Faddeev-Popov factor \cite{FaddeevPopov} can be defined as $\det Q^{\alpha}{}_{\beta}$. In the background field approach, one expands the fields $\varphi^i$ as the sum of background-fields $\bar{\varphi}^i$ and quantum fields $\eta^i$,
\begin{equation}
\varphi ^i = \bar{\varphi}^i +\eta ^i.
\end{equation}
A convenient gauge condition for $\eta ^i$ could be chosen in a practical calculation and physical results should be independent of this choice. We shall choose Landau-DeWitt gauge condition \cite{FradkinTseytlin} which has the following feature and can simplify the calculation significantly
\begin{equation}
\chi_\alpha =K_{\alpha i}[\bar{\varphi}]\eta^i=0.
\end{equation}
As aforementioned, gauge condition dependence can be reduced to parametrization dependence of the gauge field. To get rid of parametrization dependence, the field space with its metric $g_{ij}[\varphi]$ is introduced and its contribute in the effective action will cancel the above dependence. At one-loop order with Landau-DeWitt gauge, Vilkovisky-DeWitt effective action is given by
\begin{eqnarray}\label{Eq.VDaction}
& & \Gamma[\bar{\varphi}]  =  S[\bar{\varphi}]-\ln\det Q_{\alpha\beta}[\bar{\varphi}] \nonumber \\
& & +  \frac{1}{2}\lim_{\Omega\rightarrow0}\ln\det\left(\nabla^i\nabla_j S[\bar{\varphi}] + \frac{1}{2\Omega}K^{i}_{\alpha}[\bar{\varphi}] K^{\alpha}_{j}[\bar{\varphi}]\right),
\end{eqnarray}
with $\nabla_i\nabla_j S[\bar{\varphi}]=S_{,ij}[\bar{\varphi}] - \Gamma^{k}_{ij} S_{,k}[\bar{\varphi}]$. With the covariant derivative on field space, this effective action now is invariant under reparametrization of $\varphi'_i=f(\varphi_i)$. Here the Christoffel connection $\Gamma^{k}_{ij}$ is determined by $g_{ij}[\varphi]$. Note that if any other gauge condition is chosen, Eq.~({\ref{Eq.VDaction}}) will not be true and other complicated form will replace it with other $\bar{\Gamma}^{k}_{ij}$\cite{Vilkovisky,DeWitt}. It should be mentioned that it is the connection term $\Gamma^{k}_{ij} S_{,k}[\bar{\varphi}]$ that distinguishes the Vilkovisky-DeWitt's method from the traditional background-field method and cancels the gauge condition dependence.

To calculate the above effective action Eq.~(\ref{Eq.VDaction}), one can rewrite the determinant back to the functional integral
\begin{eqnarray}\label{Eq.ordertwo}
\Gamma_G &=&
\frac{1}{2} \ln\det \left\{ \nabla^i\nabla_j S[\bar{\varphi}] +\frac{1}{2\Omega}K^{i}_{\alpha}[\bar{\varphi}] K^{\alpha}_{j}[\bar{\varphi}]\right\} \nonumber \\
& = & -\ln\int\left[ d\eta\right]\,e^{-S_q}\;, \\
\Gamma_{GH} & = & -\ln\det Q_{\alpha\beta}=-\ln\int\left[ d\bar{\eta}d\eta\right] e^{-S_{GH}},
\end{eqnarray}
with $S_q  = \frac{1}{2}\eta^i\eta^j(\nabla_i\nabla_j S  +\frac{1}{2\Omega}K_{\alpha\,i}K^{\alpha}_{j})$ and
$S_{GH} = \bar{\eta}_\alpha Q^{\alpha}{}_{\beta}\eta^\beta$.
Here $\Omega\rightarrow0$ is understood to enforce the Landau-DeWitt gauge condition and all terms are evaluated at the background-field $\bar{\varphi}$. $\Gamma_{GH}$ is the ghost contribution with $\bar{\eta}_\alpha$ and $\eta^\beta$ are anti-commuting ghost fields.

{\bf Calculation of Quadratic Divergence}: We now  apply the above formalism to gravity-gauge system. For simplicity, we only consider the $U(1)$ electromagnetic theory, but the results are also true for non-abelian gauge theories. For a comparison with the results of \cite{DJToms}, we shall use the same notation. The classical action of Einstein-Maxwell theory is
\begin{equation} \label{eq:GravityL}
S =\int d^4x |g(x)|^{1/2}\left[\frac{1}{4} F_{\mu\nu}F^{\mu\nu} -\frac{2}{\kappa^2}(R-2\Lambda)\right],
\end{equation}
with $F_{\mu\nu}=\partial_{\mu}A_{\nu}-\partial_{\nu}A_{\mu}$ and $\kappa^2=32\pi G$. The cosmological constant term will contribute logarithmic divergences that will  also change $\beta$ function of gauge coupling \cite{DJToms}. We have made an independent check that the logarithmic divergence is the same as the one given in \cite{DJToms}. Here we only present the calculation of quadratic divergences. The metric $g_{ij}[\varphi]$ on the field space is chosen conventionally as follows\cite{FieldMetric}
\begin{eqnarray}\label{Eq:FieldMetric}
g_{g_{\mu\nu}(x)g_{\lambda\sigma}(x')}&=&\frac{1}{2\kappa^2}|g(x)|^{1/2}\left(g^{\mu\lambda}g^{\nu\sigma} + g^{\mu\sigma}g^{\nu\lambda}-g^{\mu\nu}g^{\lambda\sigma}\right)\delta(x,x'), \\
g_{A_{\mu}(x)A_{\nu}(x')}&=&|g(x)|^{1/2}g^{\mu\nu}(x)\delta(x,x').
\end{eqnarray}
We expand the fields, $\varphi^i=(g_{\mu\nu}, A_{\mu})$, at the background-fields, $\bar{\varphi}^i = (\delta_{\mu\nu},\bar{A}_\mu)$, as
\begin{eqnarray}
g_{\mu\nu}&=&\delta_{\mu\nu}+\kappa h_{\mu\nu}, \quad A_{\mu}=\bar{A}_\mu + a_{\mu},
\end{eqnarray}
and choose the Landau-DeWitt gauge conditions ($\omega =1$),
\begin{eqnarray}
\chi _{\lambda} &=&\frac{2}{\kappa}(\partial^\mu h_{\mu\lambda}-\frac{1}{2}\partial_\lambda h)+\omega(\bar{A}_\lambda \partial^\mu a_\mu+a^\mu\bar{F}_{\mu\lambda}), \label{Eq:GravityGaugefix}  \\
\chi & = & {}-\partial^\mu a_\mu, \label{Eq:GaugeGaugefix}
\end{eqnarray}
where $\omega$ is a parameter introduced for a comparison with the traditional background-field method. It is tempting to impose $\partial^\mu a_\mu=0$ in Eq.~(\ref{Eq:GravityGaugefix}) for simplicity, while there is subtlety in doing this and we shall discuss it elsewhere in detail \cite{TangWuL}.  Also a parameter, $v$,  is introduced for the connection terms,
\begin{equation}\label{Eq:vConnection}
S_q=\frac{1}{2}\eta^i\eta^j\left(S_{,ij}-v\Gamma^{k}_{ij}S_{,k} +\frac{1}{2\Omega}K_{\alpha\,i}K^{\alpha}_{j}\right).
\end{equation}
The gauge fixed term can be written explicitly as
\begin{equation}
S_{GF}=\frac{1}{4\Omega}\eta^i\eta^jK_{\alpha\,i}K^{\alpha}_{j}=\frac{1}{4\xi}(\chi_{\lambda})^2+\frac{1}{4\zeta}(\chi)^2,
\end{equation}
where $\xi$ and $\zeta$ are gauge fixing parameters for gravity and gauge fields, respectively.

Note that both $\omega$ and $v$ are not real gauge condition parameters, and their values are actually fixed in Landau-DeWitt gauge, $\omega =1$, $v =1$. They are introduced \cite{DJToms} just for an advantage of comparing with traditional background field method in harmonic gauge. In principle, the Vilkovisky-DeWitt formalism is applicable in any gauge condition as it has been verified to be gauge condition independent\cite{FradkinTseytlin,BV,Huggins}. While in a practical calculation, such a formalism becomes much simple in Landau-DeWitt gauge. Therefore, we will impose eventually the Landau-DeWitt gauge condition: $\omega =1$, $v =1$, $\xi \rightarrow 0$ and $\zeta \rightarrow 0$ to obtain a gauge condition independent result. Meanwhile, by taking $\omega =0$, $v =0$, $\xi = 1/\kappa^2$ and $\zeta = 1/2$, we can straightforwardly read off the results in the traditional background field method in harmonic gauge.

For a consistent check, we have reproduced the results given in \cite{DJToms} for the logarithmic divergent contributions to the $\beta$ function when cosmological constant is included. The index contraction is done with the help of FeynCalc \cite{feyncalc}. Here we only show the calculation of quadratic divergence. There are several classes of contributions. $S_q$ can be written as $S_q = S_0 + S_1 + S_2$ with the subscript denoting the order in the background gauge field $\bar{A}_\mu$ after expansion of action \cite{DJToms}.

The contributions from gravity-gauge coupling to effective action can be written as
\begin{equation}
\Gamma_{G}=\langle S_2\rangle-\frac{1}{2}\langle S_1^2\rangle,\quad \langle S_2\rangle=\langle S_{21}\rangle+\langle S_{22}\rangle,
\end{equation}
where $\langle S_{21}\rangle$ and $\langle S_{22}\rangle$ represent contributions with vertex $\bar{A}\bar{A}hh$ and $\bar{A}\bar{A}aa$, respectively. And $\langle S_1^2\rangle$ denotes contributions with vertex $\bar{A}ha$. Notice that a four pure gauge vertex appears in $\langle S_{22}\rangle$, which is tightly connected with gravity gauge fixing term in Eq.~(\ref{Eq:GravityGaugefix}) and the connection term in Eq.~(\ref{Eq:vConnection}) as it depends on $\omega$ and $v$. All the contributions involve the following quadratically divergent tensor- and scalar-type loop integrals
\begin{equation}
\mathcal{I}_{2\mu\nu} = \intp{p}\frac{p_\mu p_\nu }{p^4},\quad  \mathcal{I}_2 = \intp{p}\frac{1}{p^2} \label{Eq.Tensor+Scalar}.
\end{equation}
In general, one needs a consistent regularization to make the quadratically divergent integrals well-defined. Without involving the details of regularization schemes, one can always relate the regularized tensor-type integral with the regularized scalar-type integral via the general Lorentz structure as follows
\begin{equation}
\mathcal{I}^R_{2\mu\nu} = a_2 \delta_{\mu\nu} \mathcal{I}^{R}_2\label{Eq.TensorScalar}.
\end{equation}
Here $a_2$ may be different in different regularization schemes. However, by explicitly calculating one loop diagrams of gauge theories, it has been shown \cite{YLW} that a consistency condition with
\begin{equation}\label{eq.consistentc}
 a_2 = 1/2
\end{equation}
is required to preserve gauge invariance for $\mathcal{I}^{R}_2\neq 0$.

For completeness and more clear, let us briefly outline the proof of the consistency conditions, a detailed proof is referred to ref.\cite{YLW}. We shall work with the following general lagrangian, with $\xi$ as a gauge parameter,
\begin{eqnarray}
{\cal L} = \bar{\psi}_n (i\gamma^{\mu}D_{\mu} - m) \psi_n
   - \frac{1}{4} F^a_{\mu\nu}F_a^{\mu\nu} - \frac{1}{2\xi} (\partial^{\mu} A_{\mu}^a )^2
    + \partial^{\mu}\eta^{\ast}_a D_{\mu} \eta^a ,
\end{eqnarray}
with
 \begin{eqnarray}
  & & F_{\mu\nu}^a  =  \partial_{\mu} A_{\nu}^a - \partial_{\nu} A_{\mu}^a
  -g f_{abc}A_{\mu}^b A_{\nu}^b
  \\
  & & D_{\mu}\psi_n  = (\partial_{\mu} + ig T^a A_{\mu}^a)\psi_n
   \end{eqnarray}
where $\psi_n$, $A_{\mu}$ and $\eta$ are fermions, gauge bosons and ghost fields, respectively. $T^a$ are the generators of gauge group and $f_{abc}$ the structure function of the gauge group with $ [T^a, \ T^b ] = i f_{abc} T^c $.  Power counting shows that the most divergent Green's function is the self-energy diagram for gauge boson, which is quadratically divergent. Here we would like to present the results carried out by using the usual Feynman rules with the general $\xi$ gauge\cite{Feynmanrule}. It has been shown to be very useful to introduce the Irreducible Loop Integrals (ILIs)\cite{YLW} at one loop level as follows,
\begin{eqnarray}
& & I_{-2\alpha} = \int \frac{d^4 k}{(2\pi)^4}\ \frac{1}{(k^2 - {\cal M}^2)^{2+\alpha}} \\
& & I_{-2\alpha\ \mu\nu} = \int \frac{d^4 k}{(2\pi)^4}\
 \frac{k_{\mu}k_{\nu}}{(k^2 - {\cal M}^2)^{3 + \alpha} }\ ,  \qquad \alpha =-1, 0, 1, 2, \cdots
\end{eqnarray}
It is seen that $\alpha =-1, 0 $ denote quadratic and logarithmic divergences. ${\cal M}^2$ is a function of external momentum, masses and Feynman parameters, but it is independent of the momentum $k$.

The fermion loop contribution to the gauge self-energy diagram is given by
 \begin{eqnarray*}
\Pi_{\mu\nu}^{(f) ab}
    & = & -g^2 4N_f C_2  \delta_{ab} \  \int_{0}^{1} dx\ [\ 2 I_{2\mu\nu} (m) - I_2(m) g_{\mu\nu}\\
     &&  + 2x(1-x) (p^2 g_{\mu\nu} - p_{\mu}p_{\nu} ) I_0(m) \ ]
 \end{eqnarray*}
$p$ is the momentum of the external gauge boson and $N_f$ is the number of fermions with $tr T^aT^b = C_2 \delta_{ab}$. It can easily be seen that the first line of the above equation is quadratically divergent and violate gauge invariance, namely $p^{\mu}\Pi_{\mu\nu}^{(f) ab} \neq 0$. Only when $I_{2\mu\nu} = \frac{1}{2}g_{\mu\nu} I_2 $, i.e., $a_2=1/2$, then the gauge invariance is maintained. The gauge boson and ghost loop contributions to the gauge self-energy diagram are found to have the follow general form,
  \begin{eqnarray}
&&  \Pi_{\mu\nu}^{(g) ab} = g^2 C_1 \delta_{ab}(p^2g_{\mu\nu} - p_{\mu}p_{\nu})
  \ \int_{0}^{1} dx\ \{ \ [1 + 4x(1-x)]\ I_0\  \nonumber \\
&& +  \frac{1}{2}\lambda \ [\ \left(\ 1 + 6x(1-x)(a_0 + 2) - 3a_0 \right)
    I_{0}\ -  2x(1-x) \left(\ 1 + 12x(1-x)\ \right) p^2\ I_{-2} \ ] \nonumber \\
&& +  \frac{3}{4}\lambda^2 \ a_{-2}\ x(1-x)\ p^2\ I_{-2}\ \} \nonumber  \\
&& +   g^2 C_1 \delta_{ab}\ \int_{0}^{1} dx\ \{\ 2(\ 2I_{2\mu\nu} - I_{2}g_{\mu\nu}\ )
   +  \lambda (a_0 -1)\ p_{\mu}p_{\nu}\ x(1-x)\ p^2 \ I_{-2}\ \}
   \end{eqnarray}
where $\lambda=1-\xi$ and $f_{acd}f_{bcd} = C_1 \delta_{ab}$, and we have also used the following definitions
\begin{eqnarray}
   & & I_{0\mu\nu} = \frac{1}{4} a_0\ I_0\ g_{\mu\nu}, \qquad I_{-2\mu\nu} = \frac{1}{4} a_{-2}\ I_{-2}\ g_{\mu\nu}
\end{eqnarray}
where $I_{-2}$ and $I_{-2 \mu\nu}$ are the convergent integrals and $a_{-2} = 2/3$\cite{YLW}. Note that $\Pi_{\mu\nu}^{(g) ab}$ depends on the gauge parameter $\xi$. This is normal as Green's function can be gauge condition dependent, only the S-matrix elements are gauge condition independent. While $\Pi_{\mu\nu}^{(g) ab}$ has to satisfy the condition $p^{\mu}\Pi_{\mu\nu}^{(g) ab} = 0$ due to gauge invariance. Notice that in the last line of $\Pi_{\mu\nu}^{(g) ab}$, it concerns both quadratically and logarithmically divergent integrals and both of them can violate gauge invariance. Only with the consistency conditions $a_2=1/2$ and $a_0 = 1$, then gauge invariance is preserved.

It then needs a regularization scheme to make the divergent integrals well-defined in order to yield the consistency conditions. This is because from the naive analysis of Lorentz decomposition and tensor manipulation, namely multiplying $g^{\mu\nu}$ on both sides of the following general relation
\begin{equation}
I_{2\mu\nu } = a_2 g_{\mu\nu} I_2, \quad \to \quad  g^{\mu\nu} I_{2\mu\nu} = g^{\mu\nu} g_{\mu\nu} a_2 I_2, \quad I_2 = 4 a_2 I_2
\end{equation}
one simply obtains $a_2 =1/4$, which will destroy the gauge invariance. The reason is that for divergent integrals which are in general not well defined without using proper regularization scheme, the tensor manipulation and integration do not commute with each other, so that the resulting  consequence for divergent integration is in general not consistent. Thus one has to perform a convergent integration in order to obtain a consistent result. To see that, let us consider the time component on both sides of the above equation, i.e.,
\begin{equation}
I_{2\ 00 } = a_2 g_{00} I_2
\end{equation}
when rotating the four-dimensional energy momentum into Euclidean space via a Wick rotation, and integrating over the zero component of energy momentum $k_0$ on both sides, we have
\begin{eqnarray}
 I_{2} & = & -i\int \frac{d^4 k}{(2\pi)^4}\ \frac{1}{k^2 + {\cal M}^2} = -i\int \frac{d^3 k}{(2\pi)^4}\ \int dk_0 \frac{1}{k^2_0 + \bf{k}^2 + {\cal M}^2}  \nonumber \\
 & = & -i\int \frac{d^3 k}{(2\pi)^4}\ 2\frac{1}{\sqrt{\bf{k}^2 + {\cal M}^2}}
 \tan^{-1} \left( k_0/\sqrt{\bf{k}^2 + {\cal M}^2}\right) |_{k_0=0}^{k_0=\infty} \nonumber \\
 & = & -i\int \frac{d^3 k}{(2\pi)^3}\ \frac{1}{2\sqrt{\bf{k}^2 + {\cal M}^2}}
\end{eqnarray}
for the right-hand side, and
\begin{eqnarray}
 I_{2\ 00} & = & -i\int \frac{d^4 k}{(2\pi)^4}\ \frac{k_0^2}{(k^2 + {\cal M}^2)^2} = -i\int \frac{d^3 k}{(2\pi)^4}\ \int dk_0 \frac{k_0^2}{(k^2_0 + \bf{k}^2 + {\cal M}^2)^2} \nonumber \\
 & = & -i\int \frac{d^3 k}{(2\pi)^4}\ \int dk_0 \left( \frac{1}{k^2_0 + \bf{k}^2 + {\cal M}^2} -  \frac{\bf{k}^2 + {\cal M}^2}{(k^2_0 + \bf{k}^2 + {\cal M}^2)^2} \right) \nonumber \\
 & = &
-i\int \frac{d^3 k}{(2\pi)^4}\ \int dk_0 \left( \frac{1}{k^2_0 + \bf{k}^2 + {\cal M}^2} -  \frac{1}{2}  \frac{1}{k^2_0 + \bf{k}^2 + {\cal M}^2}\right) -\frac{k_0}{k^2_0 + \bf{k}^2 + {\cal M}^2}|_{k_0=0}^{k_0=\infty}  \nonumber \\
 & = & \frac{-i}{2}\int \frac{d^3 k}{(2\pi)^4}\ 2\frac{1}{\sqrt{\bf{k}^2 + {\cal M}^2}}
 \tan^{-1} \left( k_0/\sqrt{\bf{k}^2 + {\cal M}^2}\right) |_{k_0=0}^{k_0=\infty} \nonumber \\
 & = & \frac{-i}{2}\int \frac{d^3 k}{(2\pi)^3}\ \frac{1}{2\sqrt{\bf{k}^2 + {\cal M}^2}}  = \frac{1}{2} g_{00} I_2
\end{eqnarray}
for the left-hand side. When comparing the above results with both left and right hand sides, we arrive at $a_2=1/2$ which agrees with the consistency condition. As the above integration over the zero component of momentum $k_0$ is convergent, which is safe for any algebraic manipulation, we then come to the conclusion that the general relation for tensor-type and scalar-type quadratically divergent integrals with $a_2=1/2$ must be the exact consistency condition.

From the above demonstration, we are convinced to work out a concrete regularization scheme which can realize the consistency conditions for both quadratical and logarithmic divergent integrals. It has explicitly been proved\cite{YLW} that the loop regularization scheme satisfies the consistency conditions $a_2=1/2$ (Eq.~(\ref{eq.consistentc})) and $a_0=1$, while the naive cut-off scheme does not as it gives $a_2 = 1/4$. The dimensional regularization scheme also leads to $a_2 =1/2$ for ${\cal M}^2 \neq 0$ and $a_0 =1$, but the resulting $\mathcal{I}^{R}_2$ is suppressed to be a logarithmic divergence with multiplying by the mass scale ${\cal M}^2$ and is found to be $\mathcal{I}^{R}_2=0$ for ${\cal M}^2 =0$, as it does not preserve the divergent behavior of the loop integrals.

Taking the general relation given in Eq. (\ref{Eq.TensorScalar}) without applying any regularization scheme, we find that the quadratically divergent parts of the effective action can be written as the following general form
\begin{eqnarray}
\langle S_{2}\rangle &= & \kappa^2 (C_{21} + C_{22})\mathcal{I}^{R}_2 \frac{1}{4} \int d^4x \bar{F}^2, \nonumber \\
\langle S_{1}^2\rangle & = & \kappa^2 C_{11}\mathcal{I}^{R}_2 \frac{1}{4} \int d^4x \bar{F}^2, \nonumber \\
C_{21} & = & \frac{1}{2}\Big( [v(1-4a_2)+6a_2](\kappa^2\xi-1)+3 \Big), \nonumber  \\
C_{22} & = & \frac{v}{8}(4a_2-1)(2\zeta-1)+\frac{\omega^2}{\kappa^2\xi}\left(( 2\zeta-1)a_2+1\right), \nonumber \\
C_{11} & = & \frac{2\omega^2}{\kappa^2\xi}([2\zeta-1]a_2+1)+2\kappa^2\xi(1-a_2)\nonumber \\
& + & 6a_2-4\omega(1-a_2).
\end{eqnarray}
Thus the total graviton's contribution to the effective action is given at one-loop order
\begin{eqnarray}
\Gamma_{G} & = & \langle S_2 \rangle - \frac{1}{2}\langle S_1^2\rangle = \kappa^2 C_G \mathcal{I}^{R}_2 \frac{1}{4} \int d^4x \bar{F}^2,
\\
C_G & = &\frac{(4a_2-1)}{8}\Big(v\left[(2\zeta -1)-4(\kappa^2\xi-1)\right] \nonumber \\
& + & 8(\kappa^2\xi-1)- 16\omega - 4\Big)+6\omega a_2.
\end{eqnarray}
One intermediate check of the formalism and our calculation is to notice that the $1/\xi$ terms in $\langle S_{2}\rangle$ and $\langle S_1^2\rangle$ cancel each other for its consistency. If they do not, then it would be inconsistent when $\xi \rightarrow 0$ in Laudau-DeWitt gauge condition. Another observation is that the above result depends on $a_2$, which characterizes the dependence of different regularization schemes.

The final piece comes from the ghost's contribution to the effective action,  which can also be written as
\begin{equation}
\Gamma_{GH}=\langle S_{GH2}\rangle-\frac{1}{2}\langle S_{GH1}^2\rangle.
\end{equation}
Their quadratically divergent contributions are found to be
\begin{eqnarray}
\langle S_{GH2}\rangle &=& -\kappa^2 \omega \mathcal{I}^{R}_2 \frac{1}{4}\int d^4x \bar{F}^2\ , \quad
\left\langle S_{GH1}^2 \right\rangle =  0,
\end{eqnarray}
which is independent of $a_2$ in Eq.~(\ref{Eq.TensorScalar}), namely, regularization independent.

Thus the total quadratically divergent one-loop gravitational contribution to the effective action is given
\begin{eqnarray}\label{Eq.EAfinal}
\Gamma &=& \frac{1}{4}\int d^4x \bar{F}^2 +  \kappa^2 C \mathcal{I}^{R}_2\frac{1}{4}\int d^4x \bar{F}^2,
\end{eqnarray}
where the constant $C$ is given by
\begin{eqnarray}
C & = & C_G - \omega = \frac{4a_2-1}{8}\Big(v\left[(2\zeta -1)-4(\kappa^2\xi-1)\right]\nonumber \\
& + &8(\kappa^2\xi-1) - 16\omega - 4\Big) + \omega(-1+6a_2).
\end{eqnarray}

To get the finite contributions, the corresponding counter-term has to be added by the renormalization of gauge field and gauge coupling constant. The renormalized gauge action can be written as
\begin{eqnarray}
S_M &=& \frac{1}{4}(1 + \delta_A) \int d^4x \bar{F}_{\mu\nu}\bar{F}^{\mu\nu},
\end{eqnarray}
where $\delta_{A}$ is determined from Eq.~(\ref{Eq.EAfinal}) via the cancelation of the quadratic divergence $\delta_A + \kappa^2 C \mathcal{I}^{R}_2 \simeq 0$, namely
\begin{equation}
 \delta_A \simeq - \kappa^2 C \mathcal{I}^{R}_2,
\end{equation}
where we have considered only the quadratically divergent part  and neglected the logarithmic divergence. In the Maxwell theory, the charge renormalization constant $Z_e$ is connected to the gauge field renormalization constant $Z_A=1+\delta_A $ with $Z_eZ^{1/2}_{A}=1$, which allows us to obtain the gravitational correction to the $\beta$ function
\begin{eqnarray}
\beta^{\kappa}_{e}= \mu \frac{\partial}{\partial\mu}e=\mu \frac{\partial}{\partial\mu} Z^{-1}_e e^{0}
         =\frac{1}{2}e\mu \frac{\partial}{\partial\mu} \delta_A.
\end{eqnarray}
For any regularization scheme which maintains the original divergent behaviour of integrals, one has $\mathcal{I}^{R}_2\neq 0$, its general form for the quadratically divergent part (in the Euclidean space) can be written as follows
\begin{equation}\label{Eq:QuadInt}
\mathcal{I}^{R}_{2}\simeq \frac{1}{16\pi^2}(M_c ^2-\mu ^2),
\end{equation}
where $M_c$ and $\mu$ are the UV energy scale and renormalization energy scale, respectively. Thus the $\beta$ function correction reads
\begin{equation}\label{Eq.Betafunc}
\beta^{\kappa}_{e} = \frac{\mu ^2}{16\pi^2} e \kappa^2 C,
\end{equation}
which indicates that there exist quadratically divergent gravitational contributions to the gauge coupling constant for $C\neq 0$. With considering the cosmological constant $\Lambda$ \cite{DJToms}, the logarithmic divergence also contributes to the $\beta$ function and the above result is extended to be
\begin{equation}\label{Eq.Betaall}
\beta^{\kappa}_{e} = \frac{\mu ^2}{16\pi^2} e \kappa^2 C -\frac{3\Lambda}{64\pi^2}e \kappa^2.
\end{equation}

To give the explicit result, let us impose the Landau-DeWitt gauge condition $v=1,\ \omega =1,\ \zeta = 0,\  \xi = 0$, and take the gauge invariance consistent condition $a_2=1/2$, we then have
\begin{equation}
\beta^{\kappa}_{e} = -\frac{9\mu ^2}{128\pi^2} e \kappa^2\ ; \  C =6a_2 - 1 - \frac{25}{8}(4a_2 - 1) = -\frac{9}{8}.
\end{equation}
This result is gauge condition independent ensured by the Vilkovisky-DeWitt formalism, although we work in Landau-DeWitt gauge condition. Gauge condition independence of Vilkovisky-DeWitt's effective action has been shown a long time ago\cite{BV, Huggins}.  Also this result is universal in any regularization scheme if it can satisfy the consistent condition $a_2=1/2$ and preserve divergent behaviour of quadratically divergent integrals, $\mathcal{I}_2 \sim M_c^2 $.

It is noted that the above result is mainly based on the gauge symmetry requirement with $a_2=1/2$ and the power counting analysis with $I_2\sim M_c^2 $, we then arrive at a conclusion that the gauge symmetry and the power counting are sufficient to yield non zero quadratic contributions to the gauge couplings from the gravitational sector. Thus we come to the general statement that the quadratically divergent gravitational contributions to the gauge coupling constant is power-law running and asymptotically free.  While the concrete result for the scalar-type quadratic integral $I_2$ may yield regularization scheme dependent result when one adopts regularization schemes which cannot appropriately treat the quadratic divergence, which will be discussed below.

{\bf Discussion}: We are now in a position to make comments on the regularization scheme dependence. In the dimensional regularization, one has $\mathcal{I}^{R}_{2} = 0$ and get $\delta_A = 0$, $\beta_e^{\kappa}=0$, and no quadratically divergent gravitational contributions. In the cut-off regularization, one has $a_2=1/4$, $C= 1/2$ and $\beta^{\kappa}_{e} = \mu ^2/(32\pi^2) e \kappa^2$, which shows that there is no asymptotic freedom when the cut-off regularization scheme is used in a gauge condition independent formalism.

Note that in any case our result is different from the recent calculations given in \cite{heatkernel} by using the heat kernel scheme within the framework of Vilkovisky-DeWitt formalism and also in \cite{he} by using ``dimensional reduction'' approach. As already noticed in ref. \cite{heatkernel} that the result obtained in ref.\cite{heatkernel} is \textit{at variance with using a momentum space cut-off} \cite{Ebert}. So the result in \cite{heatkernel} is actually not consistent with the one given in \cite{Ebert, TangWu} when it goes back to harmonic gauge in traditional background field method. Also the quartic divergence may be encountered in \cite{heatkernel}, which can destroy gauge invariance due to the quartically divergent correction to mass term, $\bar{A}_{\mu}\bar{A}^{\mu}$. In \cite{he}, the field metric $g_{ij}$ used there differs by a factor of 2 with the usual convention, and also $a_2=1/2$ is introduced in a tricky way for dealing with tensor integrals through reducing the 4-dimensional space-time to 2-dimensional space-time, which is not a systematically consistent regularization scheme for preserving gauge invariance and meanwhile maintaining divergent behavior of original integrals. If the real dimensional reduction in 4-dimension is used there, it will lead to $a_2=1/4$ and no asymptotic freedom.

As an independent check, let us revisit the traditional background field method in the harmonic gauge, which is recovered by simply taking $v=0,\ \omega =0,\ \zeta = 1/2,\  \xi = 1/\kappa^2$ in the above Vilkovisky-DeWitt formalism. As a consequence, it leads to
\begin{equation}
C = 1/2 - 2a_2.
\end{equation}
It is manifest that in the cut-off regularization, one has $a_2=1/4$, $C=0$ and $\beta^{\kappa}_{e} = 0$, which confirms the previous results given in \cite{Pietrykowski,Ebert,TangWu}. In the loop regularization scheme or any regularization schemes that preserve gauge invariance and maintain the divergent behavior of original integrals, it gives $a_2 =1/2$, we then have $C=-1/2$ and  $\beta^{\kappa}_{e} = -\mu ^2/(32\pi^2) e \kappa^2$. It reproduces our previous conclusion \cite{TangWu} that the quadratically divergent gravitational contributions to the gauge coupling constant is asymptotic free in the traditional background field or diagrammatic method with the harmonic gauge.

There are papers \cite{Donoghue,Ellis} to further address the physical meaning of the running coupling from the S-matrix elements and the mixing of operators. As stated, we only make some comments but not try to settle all things down. Noticing the potential ambiguity of $\beta$-function in effective field theory with a dimensional constant, the authors in ref.~\cite{Donoghue} creatively modified the definition of $\beta$ function used in standard renormalization group analysis and defined the running coupling in physical processes using $\lambda\varphi^4$ with dimensional regularization as an illustrative example. The calculation there showed that their new definition will not give universal running couplings, for example, off-shell and on-shell scattering process give different $\beta$ function. So it is not clean whether their approach is applicable to gauge-gravity system. In ref.~\cite{Ellis}, the authors argue that the ambiguity from the quadratic divergence stem from the undetermined coefficient of high order derivative term $\partial_{\rho}F_{\mu\nu}\partial^{\rho}F^{\mu\nu}$ and can be removed by field redefinition. It is noted that the term $\partial_{\rho}F_{\mu\nu}\partial^{\rho}F^{\mu\nu}$ can only absorb divergences like $q^2\ln(\Lambda^2)$ rather than $\Lambda^2$. Both types of divergences can appear in integrals like
\[
I_2 =\intp{p}\frac{1}{p^2-q^2},\; \quad I_{2\mu\nu} = \intp{p}\frac{p_\mu p_\nu }{(p^2-q^2)^2}
\]
when applying the loop regularization scheme\cite{YLW,LORE,LORE1} to the above integrals, we then obtain the explicit form as: $ I^R_{2\mu\nu} = \frac{1}{2} g_{\mu\nu} I_2^R $ and $I_2^R = \frac{1}{16\pi^2}[M_c^2 -q^2 - q^2(\ln M_c^2/q^2 - \gamma_E )]$ with $M_c^2\to \infty$. It should be pointed out that in dimensional regularization, the above quadratic divergence is reduced to logarithmic one, or equivalently, quadratically divergent correction like $\kappa^2\Lambda^2F_{\mu\nu}F^{\mu\nu}$ is transferred to logarithmically divergent one like $\kappa^2q^2\ln{\Lambda^2}F_{\mu\nu}F^{\mu\nu}$ which requires renormalization of higher order operator, $\partial_{\rho}F_{\mu\nu}\partial^{\rho}F^{\mu\nu}$. Then the problem of operators mixing can arise \cite{Donoghue}. As we work with loop regularization, quadratic divergence is kept and we do not see the above problem.

Note that our results also disagree with the recent calculations\cite{AF1,AF2} where the running gauge coupling was found to receive no contribution from the gravitational sector. In ref.\cite{AF1}, it shared the framework of functional renormalization group with \cite{Reuter} but obtained a different conclusion. The main point in \cite{AF1} is that it used a modified symmetry identity to criticize the choice of regularization schemes. While in our case, as mentioned, the Slavnov-Taylor identity is satisfied and other criteria, like preserving divergence behavior, has to be used. In ref. \cite{AF2}, it used the transverse feature of QED vacuum polarization in the flat space-time to argue that the quadratic divergence should be discarded since it will break gauge invariance. It is noticed that in the calculation of ref. \cite{AF2}, only the matter's correction to QED vacuum polarization break the gauge invariance, rather than the correction from the gravity. Thus its argument is not applicable to gauge-gravity system. Also, the matter's correction can be regularized with loop regularization to ensure its gauge invariance.

{\bf Conclusion}: In conclusion, we have considered the quadratically divergent gravitational corrections to the running of gauge couplings in the gauge condition independent Vilkovisky-DeWitt formalism deduced in the framework of background field method. We restrict our discussion in the framework of standard renormalization group analysis. With the consistency condition between the regularized tensor-type and scalar-type quadratically divergent integrals due to the gauge invariance requirement, we have obtained a gauge condition independent result for the $\beta$ function. The result is applicable to any regularization schemes that can satisfy a requirement for preserving both gauge invariance and original divergent behavior of integrals. The loop regularization method has been found to satisfy such a requirement and to provide a systematic and consistent approach for applying to the calculations of gravitational contributions to gauge couplings.

\begin{acknowledgments}
The authors would like to thank Jianwei Cui and Da Huang for useful discussions. This work was supported in part by the National Science Foundation of China (NSFC) under the grant \# 10821504, 10975170 and the key Project of Knowledge Innovation Program (PKIP) of Chinese Academy of Science.
\end{acknowledgments}


\begin{thebibliography}{999}
\bibitem{convention}
  Note that the field metric $g_{ij}$ used by two groups differ by a factor of 2, and also different regularization schemes are adopted by two groups.
\end{thebibliography}
\end{document}